\documentstyle[aps,preprint]{revtex}
\begin{document}
\draft
\newcommand{\bgeq}{\begin{equation}}
\newcommand{\eneq}{\end{equation}}
\newcommand{\bgeqa}{\begin{eqnarray}}
\newcommand{\eneqa}{\end{eqnarray}}
\newcommand{\bgit}{\begin{itemize}}
\newcommand{\enit}{\end{itemize}}
\newcommand{\bgen}{\begin{enumerate}}
\newcommand{\enen}{\end{enumerate}}
\renewcommand{\theequation}{\arabic{equation}}
\newcommand{\neweqn}[1]{\renewcommand{\theequation}{\arabic{equation}}
                       \setcounter{equation}{#1}}
\newcommand{\neweqs}[1]{\renewcommand{\theequation}
                                     {\arabic{section}.\arabic{equation}}
                       \setcounter{equation}{#1}}
\newcommand{\neweqc}[1]{\renewcommand{\theequation}
                                     {\arabic{chapter}.\arabic{equation}}
                       \setcounter{equation}{#1}}
\newcommand{\eqreset}{\setcounter{equation}{0}}
\renewcommand{\thefootnote}{\fnsymbol{footnote}}

\title{Spin Gap in Two-Dimensional Heisenberg Model for CaV$_4$O$_9$}
\author{Nobuyuki {\sc Katoh} and Masatoshi {\sc Imada}}
\address{\it Institute for Solid State Physics, University of Tokyo,\\
7-22-1, Roppongi, Minato-ku Tokyo 106, Japan}

\date{\today}
\maketitle

\begin{abstract}
{\bf ABSTRACT:}
  We investigate the mechanism of spin gap formation in a two-dimensional
model relevant to Mott insulators such as CaV$_4$O$_9$. From the
perturbation expansion and quantum Monte Carlo calculations, the origin of
the spin gap is ascribed to the four-site plaquette singlet in contrast to
the dimer gap established in the generalized dimerized Heisenberg model.

\end{abstract}

\hspace{7cm}(To be published in J.Phys.Soc.Jpn.)

\hspace{7cm}(e-mail address: katoh@falcon.issp.u-tokyo.ac.jp)

\newpage

     In previous papers,
\cite{NKatohJPSJ62-3728,NKatohJPSJ63-4529,NKatohJPSJ64-1437}
we have investigated the mechanism of spin gap formation in a class of
two-dimensional generalized antiferromagnetic Heisenberg (AFH) models with
dimerization.
{}From this analysis, the origin of the spin gap in low-dimensional systems
such as the spin-Peierls system, the ladder system and the Haldane system
has been identified as the dimer gap in a unified way.

     Recently, Taniguchi {\it et al.} found that the temperature dependence
of the uniform magnetic susceptibility and the nuclear magnetic relaxation
time $T_1$ of CaV$_4$O$_9$ indicate the existence of the spin gap.
\hspace{-0.5cm}
\cite{STaniguchiPre}
In this paper, we investigate the possible mechanism of the spin gap in
CaV$_4$O$_9$.

     The crystal structure of VO$_5$ pyramid layer in CaV$_4$O$_9$ is
shown in Fig.
\ref{fig-1}(a).
Since the valence of V atom is 4+, V atom can be treated as the localized spin
with $S=1/2$. In the case of copper oxide systems such as high-$T_c$
materials and Sr$_{n-1}$Cu$_{n+1}$O$_{2n}$,
\cite{ZHiroiSSC95-230}
the electrons are localized in the $d_{\gamma}$ orbitals on Cu$^{2+}$ sites.
However in this system, the Fermi level lies in the $d_{\epsilon}$ orbital
and hence the localized spins should mainly come from electrons in
the $d_{\epsilon}$ orbitals on V$^{4+}$ sites.
The degree of the orbital degeneracy with three-fold
degenerate $d_{\epsilon}$ orbitals has not been clarified yet,
because the anisotropic crystal field due to possible Jahn-Teller
distortion has not been estimated precisely.
In any case, the antiferromagnetic exchange coupling is expected between the
adjacent $d_{\epsilon}$ orbitals on the nearest-neighbor V atoms.
We also expect that the exchange coupling between the nearest neighbors is
dominant due to the $d_{\epsilon}$ character.
Based on these facts, the $S=1/2$ AFH model with the network shown in Fig.
\ref{fig-1}(b)
is a good starting point to discuss the Mott insulating phase.
The Hamiltonian is written as
\bgeq
{\cal H}=J\sum_{\langle i,j \rangle}
\mbox{\boldmath$S$}_i \mbox{\boldmath$\cdot S$}_j,
\eneq
where $\langle i,j \rangle$ denotes the nearest-neighbor bonds.

     We first calculate the spin gap by quantum Monte Carlo (QMC)
simulation to check whether our model is relevant or not.
Here the spin gap $\Delta _{\mbox{\small s}}$ is defined
as the energy difference between the lowest state with
$S^{\mbox{\tiny z}}_{\mbox{\tiny total}}=
\sum_i S^{\mbox{\tiny z}}_{i}=1$ and
the singlet ground state.
The algorithm of the QMC simulation is
based on the world-line method combined with the Suzuki-Trotter method
\cite{MSuzukiProgTheorPhys56-1454}
and the checker-board decomposition.
\cite{JEHirschPRL47-1628,JDRegerPRB37-5978}
The temperature $T$ scaled by $J$ is taken as 0.1,
which is sufficiently low to observe the ground state properties for the
present purpose.
The Trotter numbers $N_{\mbox{\tiny T}}$ are 40, 60, 80 and 100.
The number of QMC steps is $(1 \sim 5) \times 10^7$
for each Trotter number.
The extrapolation formula used to obtain $\Delta _{\mbox{\small s}}$
from the finite $N_{\mbox{\tiny T}}$ is
given by $\Delta _{\mbox{\small s}} (L)=
\Delta _{\mbox{\small s}} (\infty) + A/N_{\mbox{\tiny T}}^2$.
The lattices we have calculated have sizes
$1 \times 1$, $2 \times 2$, $3 \times 3$ and $4 \times 4$
cells under the periodic boundary condition, where a cell includes 20 sites.
The value of the spin gap extrapolated to the
thermodynamic limit is estimated to be $(0.11 \pm 0.03)J$ with
the form of the fitting function $\Delta _{\mbox{\small s}} +A/L^2$,
as is shown in Fig.
\ref{fig-2}.

We also calculate the temperature dependence of the uniform susceptibility
for $2 \times 2$ cells, which is shown in Fig.
\ref{fig-3}.
The susceptibility in the thermodynamic limit should be close to this
result, since the the spin gap for $2 \times 2$ cells is close to the
value in the thermodynamic limit.
As a reference, the susceptibility of the conventional square-lattice
AFH model on $8 \times 8$ lattice is also shown in Fig.
\ref{fig-3},
which may describe the thermodynamic properties within statistical errors.
\cite{YOkabePRL61-2971}
This also shows the existence of the spin gap in our model.

     Next we investigate the origin of the spin gap.
The four-site plaquette is treated as the unperturbed Hamiltonian
while the other bonds are taken as the perturbation.
As is shown in Fig.
\ref{fig-4},
the strength of the spin exchange couplings
in the four-site plaquette is taken as $J$,
while the strength of the others is $J'$.
The ground state of the unperturbed Hamiltonian
is the product state of singlets on the four-site plaquettes,
in other words, a resonating valence bond state in each plaquette.
The first excited states consist of the triplet state on one of the plaquettes
and the singlet on the others. The degeneracy is lifted by the first-order
perturbation expansion (PE) through the hopping of the triplet due to the
translational symmetry.
Then the energy of the triplet state has a wave-number dependence.
We calculate the energy of the ground state and triplet states using
the second-order PE. Then the ground state energy per site
$\epsilon _{\mbox{\tiny G}}^{(2)}$ and
the energy difference between the ground state and the triplet state
$\Delta ^{(2)}_{\mbox{\small s}}(\mbox{\boldmath $k$})$ are obtained as
\bgeq
      \epsilon _{\mbox{\tiny G}}^{(2)}
      = -\frac{1}{2}J[1+\frac{43}{576}\frac{J'^2}{J^2}+ O((\frac{J'}{J})^3)],
                      \label{ref0.5}
\eneq
\bgeqa
\lefteqn{
      \Delta ^{(2)}_{\mbox{\small s}} (\mbox{\boldmath $k$})
     = J
      +\frac{1}{3}J'( \cos k_x a + \cos k_y a )
      -\frac{47}{864} \frac{J'^2}{J}
        }
                      \nonumber\\
      & &+\frac{1}{54}\frac{J'^2}{J} ( \cos ^2 k_x a + \cos ^2 k_y a )
         -\frac{1}{9} \frac{J'^2}{J} \cos k_x a \cos k_y a,
                      \label{ref1}
\eneqa
where $a$ is the lattice constant between plaquettes, as is shown in Fig.
\ref{fig-4}.
The dispersion of the triplet excitation
$\Delta _{\mbox{\small s}}^{(2)} (\mbox{\boldmath $k$})$ is shown in Fig.
\ref{fig-4.5}.
The lowest triplet excitation is located at $\mbox{\bf k} = (\pi,\pi)$. An
interesting point is that the spin gap survives $(0.205J)$ even if
the value of $J'$ is equal to $J$.
On the other hand,
if we take the unperturbed Hamiltonian as $J'$ bonds with $J/J'$
taken as the perturbation,
the ground state energy per site given by the second-order PE
$\epsilon '^{(2)} _{\mbox{\tiny G}} $
is obtained as
\bgeq
     \epsilon '^{(2)}  _{\mbox{\tiny G}}
            = -\frac{3}{8}J'(1+\frac{1}{4}\frac{J^2}{J'^2}).
                      \label{ref1.3}
\eneq
When $J'$ is equal to $J$, $\epsilon _{\mbox{\tiny G}} ^{(2)} = -0.537J$
obtained from eq.
(\ref{ref0.5}) is lower than
$\epsilon '^{(2)} _{\mbox{\tiny G}} = -0.469J$ in eq.
(\ref{ref1.3}).
The dispersion of the triplet state given by
the first-order PE in terms of $J/J'$ is described as
\bgeq
      \Delta'^{(1)} _{\mbox{\small s}} (\mbox{\boldmath $k$}) = J'-
      \frac{1}{2}J( \cos k_x a + \cos k_y a ).
                      \label{ref1.5}
\eneq
The perturbation breaks down and the spin gap closes at $J/J'=1$ even
in the first-order PE.

     The above results imply that the origin of the spin gap is basically the
four-site plaquette singlet state rather than the dimer singlet.
The spin gap in CaV$_4$O$_9$ has been estimated
from the susceptibility as $\Delta_{\mbox{\small s}} \sim 100$K,
\cite{STaniguchiPre}
whereas the gap amplitude obtained from the QMC is
$\Delta_{\mbox{\small s}} /J \sim 0.11$.
Although at present we do not have available data for $J$ in
CaV$_4$O$_9$, one may argue that the calculated value of the spin gap
appears to be smaller than the observed value
because the exchange interaction in vanadium oxides is expected to be
at most several hundred K. One possible origin of the gap enhancement
is the frustration effect arising from the next-nearest-neighbor exchange
coupling. It is also noted that the effect of the orbital degeneracy
and the orbital correlation effect could be important for a quantitative
estimation of the gap.

     In order to investigate the mechanism of the spin gap in more detail
from the viewpoint of the four-site plaquette singlet formation,
we investigate the one-dimensional analog of this model.
The lattice structure is shown in Fig.
\ref{fig-5}(a).
We calculate the spin gap by the exact diagonalization (ED) method.
The size dependence of the spin gap is shown in Fig.
\ref{fig-6}.
After the extrapolation to the thermodynamic limit, we estimate the spin gap
as $\Delta_{\mbox{\small s}} \sim 0.60J$,
which is larger than the value of the spin gap in the ladder model.
\cite{TBarnesPRB47-3196,MTroyerPRB50-13515}
The fitting function is the same as the one in the previous analysis.
In the second-order PE, the triplet state energy is obtained as
\bgeq
      \Delta _{\mbox{\small s}} (k) = J
      +\frac{1}{3}J'\cos k_x a
      -\frac{31}{1728} \frac{J'^2}{J}
      +\frac{1}{108}\frac{J'^2}{J} \cos 2 k_x a.
                      \label{ref2}
\eneq
When $J'$ is equal to $J$, the value of the spin gap is
$\Delta _{\mbox{\small s}} = 0.658J$ at $k=\pi$,
which is close to the estimation from the ED.
These results in the one-dimensional model show
that the four-site plaquette singlet is
a good starting point for discussing the
ground state in a class of lattices constructed from the four-site plaquettes
connected by a small number of bonds. This one-dimensional model
may be not a toy model but a relevant model in some transition metal oxide
compounds if a lattice structure such as that in Fig.
\ref{fig-5}(b) is realized.

     In summary, we have studied the mechanism of the spin gap in
CaV$_4$O$_9$.
{}From the QMC as well as the PE,
the origin of the spin gap formation is found to be the
four-site plaquette singlets.

     One of the authors (M.I.) thanks M. Sato for illuminating discussions
on their experimental results.
One of the authors (N.K.) thanks K. Ueda, T. Tohyama and
N. Furukawa for stimulating disccussions and useful comments.
We have used a part of the codes for the Lanczos calculation provided by
H.Nishimori in TITPACK Ver.2. A part of the computation has been
carried out on VPP500 at the Supercomputer Center of the Institute for
Solid State Physics, Univ. of Tokyo. This work is financially supported by a
Grant-in-Aid for Scientific Research  on Priority Areas "Anomalous Metallic
State near the Mott Insulator" and "Novel Electronic States in Molecular
Conductors".


\begin{figure}

\vspace*{5mm}
\caption{
(a)The lattice structure of VO$_5$ pyramid layer in CaV$_4$O$_9$. Open
circles represent V sites, while solid circles show oxygen sites.
(b)The network of the $S=1/2$ AFH model. In the QMC calculation, we take
20 sites as a cell as shown by the wavy-line square.
}
\label{fig-1}

\vspace*{5mm}
\caption{
Size dependence of the spin gap in our model.
}
\label{fig-2}

\vspace*{5mm}
\caption{
Temperature dependence of the uniform magnetic susceptibility for
$2 \times 2$ cells.
Solid circles represent the susceptibility of our model
for $2 \times 2$ cells,
while open circles show the susceptibility
for the conventional square lattice AFH model
for $8 \times 8$ lattices.
}
\label{fig-3}

\vspace*{5mm}
\caption{
The lattice structure in the perturbation calculation.
Bold lines represent the bonds with $J$ and thin lines those with $J'$.
The distance between the nearest-neighbor four-site plaquettes is $a$.
}
\label{fig-4}

\vspace*{5mm}
\caption{
The wave-number dependence of the triplet excitation energy in the
second-order perturbation calculation.
The inset shows the route of the wave vector taken for the abscissa.
}
\label{fig-4.5}

\vspace*{5mm}
\caption{
(a)The one-dimensional four-site plaquette model.
The definitions of the symbols are the same as in Fig.1(b).
(b)The lattice structure of a possible quasi-one-dimensional
four-site-plaquette system of transition metal oxides.
}
\label{fig-5}

\vspace*{5mm}
\caption{
Size dependence of the spin gap in the one-dimensional model. The sizes
we have calculated are 8,16 and 24 sites.
}
\label{fig-6}

\end{figure}


\begin{references}

\bibitem{NKatohJPSJ62-3728}N.\ Katoh and M.\ Imada:
                 \ J.Phys.Soc.Jpn. {\bf 62} (1993) 3728

\bibitem{NKatohJPSJ63-4529}N.\ Katoh and M.\ Imada:
                 \ J.Phys.Soc.Jpn. {\bf 63} (1994) 4529

\bibitem{NKatohJPSJ64-1437}N.\ Katoh and M.\ Imada:
                 \ J.Phys.Soc.Jpn. {\bf 64} (1995) 1437

\bibitem{STaniguchiPre}S.\ Taniguchi, T.\ Nishikawa, Y.\ Yasui, Y.\ Kobayashi,
                 M.\ Sato, T.\ Nishioka, M.\ Kotani and K.\ Sano:
                 \ J.Phys.Soc.Jpn. {\bf 64} (1995) 2758

\bibitem{ZHiroiSSC95-230}Z.\ Hiroi, M.\ Azuma, M.\ Takano and Y.\ Bando:
                 \ J. Solid State Chem. {\bf 95} (1991) 230

\bibitem{MSuzukiProgTheorPhys56-1454}M.\ Suzuki:
                 \ Prog.Theor.Phys. {\bf 56} (1976) 1454

\bibitem{JEHirschPRL47-1628}J.\ E.\ Hirsch, R.\ L.\ Sugar, D.\ J.\ Scalapino
                 and R.\ Blankenbecler:
                 \ Phys.Rev.B {\bf 26} (1982) 5033

\bibitem{JDRegerPRB37-5978}J.\ D.\ Reger and A.\ P.\ Young:
                 \ Phys.Rev.B {\bf 37} (1988) 5978

\bibitem{YOkabePRL61-2971}Y.\ Okabe, M.\ Kikuchi and A.\ D.\ S.\ Nagi:
                 \ Phys.Rev.Lett. {\bf 61} (1988) 2971

\bibitem{TBarnesPRB47-3196}T.\ Barnes, E.\ Dagotto, J.\ Riera and
                 E.\ S.\ Swanson:
                 \ Phys.Rev.B {\bf 47} (1993) 3196

\bibitem{MTroyerPRB50-13515}M.\ Troyer, H.\ Tsunetsugu and D.\ W\"urtz:
                 \ Phys.Rev.B {\bf 50} (1994) 13515


\end{references}
\end{document}